\documentclass[preprint,aps,superscriptaddress] {revtex4}
\usepackage{amsmath,amssymb}
\usepackage{graphics}
\usepackage{graphicx}
\usepackage{epsfig}
\begin{document}
\title{Vibrational Resonance in the Morse Oscillator
}
\author{K.~Abirami}
\email{nk.abirami@gmail.com}
\author{S.~Rajasekar}
\email{rajasekar@cnld.bdu.ac.in}
\affiliation{Centre for Nonlinear Dynamics, School  of Physics, Bharathidasan University, Tiruchirapalli 620 024, Tamilnadu}

\author{M A F Sanjuan}
\email{miguel.sanjuan@urjc.es}
\affiliation{Nonlinear Dynamics, Chaos and Complex Systems Group, Departamento de F\'{i}sica, \\ Universidad Rey Juan Carlos, Tulip\'{a}n s/n, 28933 M\'{o}stoles, Madrid, Spain}

\begin{abstract}
We investigate the occurrence of vibrational resonance in both classical and quantum mechanical Morse oscillators driven by a biharmonic force.  The biharmonic force consists of two forces of widely different frequencies $\omega$ and $\Omega$ with $\Omega \gg \omega$.  In the damped and biharmonically driven classical Morse oscillator applying a theoretical approach we obtain an analytical expression for the response amplitude at the low-frequency $\omega$. We identify the conditions on the parameters for the occurrence of the resonance. The system shows only one resonance and moreover at resonance the response amplitude is $1/(d\omega)$ where $d$ is the coefficient of linear damping. When the amplitude of the high-frequency force is varied after resonance the response amplitude does not decay to zero but approaches a nonzero limiting value. We have  observed that vibrational resonance occurs when the sinusoidal force is replaced by a square-wave force.  We also report the occurrence of resonance and anti-resonance of transition probability of quantum mechanical  Morse oscillator in the presence of the biharmonic external field.

 \vskip 10pt

\vskip 20pt
\noindent To appear in Pramana-Journal of Physics(in press, 2013)
\end{abstract}
\maketitle
\section{Introduction}
\label{sec1}
Amplification and detection of weak signals are important in many branches of science. During the past two decades, a great deal of interest has been focused on signal processing in nonlinear systems. There are few interesting ways of enhancing the response of a nonlinear system to a weak signal. In stochastic resonance an optimum weak noise amplifies the response of a nonlinear system making use of the bistability of the system [\ref{r1},\ref{r2}]. In another approach the noise is replaced by a high-frequency periodic signal. That is, the system is essentially driven by two periodic signals of widely different frequencies, say, $\omega$ and $\Omega$ with $\Omega\gg\omega$. When the amplitude of the high-frequency external force is varied the oscillation amplitude of the system at the low-frequency $\omega$ exhibits resonance. This high-frequency force induced resonance is termed as vibrational resonance [\ref{r3}-\ref{r5}]. One-way or unidirectional coupling can also be able to improve the performance of coupled systems [\ref{r6}-\ref{r8}]. Similarly, the same phenomenon can be also observed when besides a noise or a high-frequency signal, a chaotic signal is used to perturb the system [\ref{r9}-\ref{r10}].

The study of vibrational resonance in the presence of a biharmonic force has received a considerable interest in recent years after the seminal paper by Landa and McClintock [\ref{r3}]. The occurrence of vibrational resonance has been analyzed in systems with monostable [\ref{r11}], bistable [\ref{r3}-\ref{r5}], spatially periodic states [\ref{r12}], excitable systems [\ref{r13},\ref{r14}], coupled oscillators [\ref{r15}] and small world networks [\ref{r14},\ref{r16},\ref{r17}]. Experimental evidence of vibrational resonance in vertical cavity laser system [\ref{r18}-\ref{r20}] and in an electronic circuit [\ref{r21}] has also been reported. Frequency-resonance-enhanced vibrational [\ref{r22}] and undamped signal propagation in one-way coupled oscillators [\ref{r23}] and maps [\ref{r24}] assisted by biharmonic force are found to occur as well.

It is important to investigate the vibrational resonance in different kinds of systems and bring out the various features of it and the influence of characteristics of the potential of the system on it. In this connection we point out that (i) one or more resonances in monostable systems [\ref{r11}], (ii) a sequence of resonance peaks in a spatially periodic potential system [\ref{r12}], (iii) additional resonances due to asymmetry in the potential [\ref{r25}], periodic and quasiperiodic occurrence of resonance peaks in time-delayed feedback systems [\ref{r26}-\ref{r29}] and large amplitude of vibrational resonance at the dynamic bifurcation point [\ref{r30}] have been reported.

\begin{figure}[t]
\begin{center}
\includegraphics[width=0.5\linewidth]{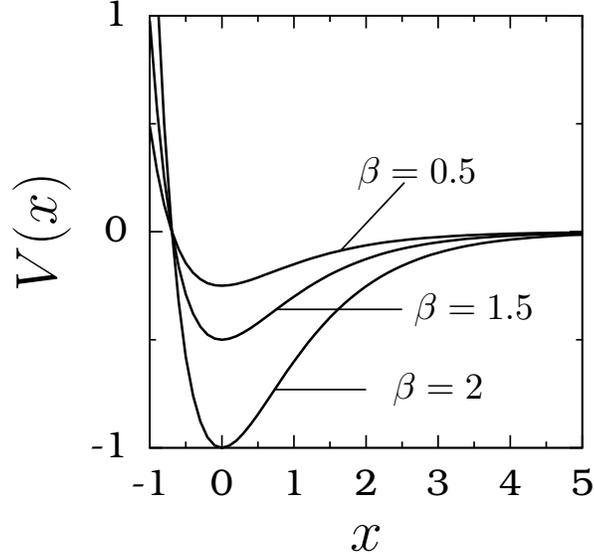}
\end{center}
\caption{The shape of the Morse potential for three values of $\beta$.}
\label{f1}
\end{figure}

Motivated from some of the previous results, in the present paper we report our investigation on the vibrational resonance in the Morse oscillator. The potential of the Morse oscillator is
\begin{equation}
\label{e1}
  V(x) = \frac{1}{2} \beta{\mathrm{e}}^{-x}
             \left( {\mathrm{e}}^{-x}-2 \right),
\end{equation}
where $\beta$ is a constant parameter representing the dissociation energy. Figure \ref{f1} depicts the form of the potential for a few values of $\beta$.  The potential is nonpolynomial and $V(x)\rightarrow\infty$ as $x\rightarrow-\infty$ while it becomes $0$ in the limit of $x\rightarrow\infty$. It has one local minimum at $x=0$ and the depth of the potential is $\beta/2$. The Morse oscillator was introduced as a useful model for the interatomic potential and fitting the vibrational spectra of diatomic molecules. It is also used to describe the photo-dissociation of molecules, multi-photon excitation of the diatomic molecules in a dense medium or in a gaseous cell under high-pressure and pumping of a local model of a polyatomic molecule by an infrared laser [\ref{r31}-\ref{r34}]. The existence and bifurcations of periodic orbits have been studied in detail [\ref{r35},\ref{r36}].

The equation of motion of the damped and biharmonically driven classical Morse oscillator is given by
\begin{equation}
\label{e2}
  \ddot{x} + d \dot{x} + \beta{\mathrm{e}}^{-x}
        \left( 1 - {\mathrm{e}}^{-x} \right)
    = f \cos \omega t + g \cos \Omega t, 
         \quad \Omega\gg\omega.
\end{equation}
Because of the difference in the time scales of the two periodic forces the motion of the system (\ref{e2}) contains both a slow component $X(t)$ with period $T=2\pi/\omega$ and a fast varying component $\psi(t,~\tau=\Omega t)$ with period $2\pi/\Omega$. That is, we assume that $x(t)=X(t)+\psi(t,\tau)$. When the amplitude $g$ or frequency $\Omega$ is varied the response of the system (\ref{e2}) at the frequency $\omega$ can exhibit resonance at a particular value of the control parameter $g$ or $\Omega$. To analyse this resonance phenomenon and the influence of the shape of the potential on the resonance we obtain an equation of motion for the slow variable applying a perturbation theory. From the solution of the linearized version of the equation of motion about its equilibrium point, we find the analytical expression for the response amplitude $Q$ which is the ratio of the amplitude of oscillation of the output of the system at the frequency $\omega$ and the amplitude $f$ of the input signal. From the expression of $Q$, we extract various features of vibrational resonance in the Morse oscillator. Particularly, we determine the value of $g$ at which resonance occurs, the maximum value of $Q$ at resonance and the limiting value of $Q$. We confirm all the theoretical predictions through numerical simulation. The theoretical treatment used for the analysis of the vibrational resonance with the periodic force $f\cos\omega t +g\cos\Omega t$ can be applied for other types of biharmonic forces. In particular, we illustrate this for a square-wave form of low-frequency and high-frequency forces.   Next, we consider the quantum mechanical Morse oscillator subjected to the biharmonic external field.  Applying a perturbation theory we obtain an analytical expression for the first-order transition probability $P_{fi}$ for a transition from an $i$th quantum state to an $f$th quantum state in time $T$ caused by the applied external field.  We analyse the influence of the high-frequency field on $P_{fi}$ and show the occurrence of resonance and anti-resonance. 
\section{Classical Morse oscillator}
In this section we consider the classical Morse oscillator described by the equation of motion (\ref{e2}) and analyse the occurrence of vibrational resonance.

\subsection{Theoretical approach}
To find an approximate solution of Eq.~(\ref{e2}), we make use of the method of separation of variables by assuming $x=X(t)+\psi(t,~\tau=\Omega t)$, where $X$ and $\psi$ are slow and fast variables respectively. Substituting $x=X+\psi$ in Eq.~(\ref{e2}) and adding and subtracting the terms $\beta\langle{\mathrm{e}}^{-\psi}\rangle {\mathrm{e}}^{-x}$ and $\beta\langle {\mathrm{e}}^{-2\psi}\rangle {\mathrm{e}}^{-2x}$ where $\langle u\rangle =(1/2\pi)\int_0^{2\pi}u~d\tau,~\tau=\Omega t$ we obtain
\begin{eqnarray}
 \label{e3}
 \ddot{X} + d \dot{X} + \beta{\mathrm{e}}^{-X}
       \left( \langle {\mathrm{e}}^{-\psi}\rangle 
        -\langle {\mathrm{e}}^{-2\psi} \rangle 
         {\mathrm{e}}^{-X} \right) 
   & = & f \cos \omega t , \\
 \label{e4} 
 \ddot{\psi} + d \dot{\psi} + \beta{\mathrm{e}}^{-X}
     \left( {\mathrm{e}}^{-\psi} - \langle
      {\mathrm{e}}^{-\psi}\rangle \right) & & \nonumber \\
         - \beta{\mathrm{e}}^{-2X} \left( 
          {\mathrm{e}}^{-2\psi} - \langle {\mathrm{e}}^{-2\psi}
            \rangle \right) 
   & = &  g \cos \Omega t .
\end{eqnarray}
Because, $\psi$ is a rapidly changing variable with fast time $\tau$ we approximate Eq.~(\ref{e4}) as $\ddot \psi=g\cos\Omega t$. The solution of this equation is $\psi=\mu\cos\Omega t$ where $\mu=g/\Omega^2$. This solution gives 
\begin{subequations}
\label{e5}
\begin{eqnarray}
  \langle {\mathrm{e}}^{-\psi}\rangle  
  & = &  \frac{1}{2}\int_0^{2\pi}{\mathrm{e}}^{\mu\cos\tau}~d
           \tau=I_0(\mu), \\
  \langle {\mathrm{e}}^{-2\psi}\rangle 
  & = & I_0(2\mu), 
\end{eqnarray}
\end{subequations}
where $I_0(z)$ is the zeroth-order modified Bessel function [\ref{r37}]. Now, Eq.~(\ref{e3}) becomes
\begin{equation}
\label{e6}
   \ddot{X} + d\dot{X} + \beta{\mathrm{e}}^{-X}
    \left( I_0(\mu) - I_0(2\mu){\mathrm{e}}^{-X} \right)
    = f \cos\omega t. 
\end{equation}
Equation (\ref{e6}) can be treated as the equation of motion of a particle experiencing the external periodic force $f\cos\omega t$ and linear friction force in the effective potential
\begin{equation}
\label{e7}
  V_{\mathrm{eff}} = \frac{1}{2} \beta{\mathrm{e}}^{-X}
      \left[ I_0(2\mu){\mathrm{e}}^{-X}-2I_0(\mu) \right] .
\end{equation}
In addition to the parameter $\beta$, the effective potential depends on the parameters $g$ and $\Omega$. The dependence of $V_{\mathrm{eff}}$ on $g$ and $\Omega$ is in the form of a zeroth-order modified Bessel function.

\begin{figure}[t]
\begin{center}
\includegraphics[width=0.85\linewidth]{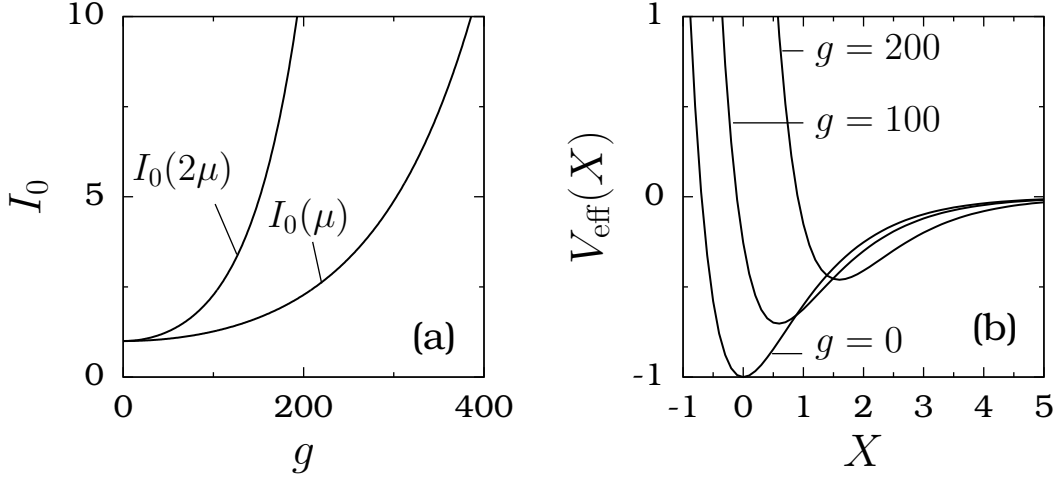}
\end{center}
\caption{(a) Dependence of a zeroth-order modified Bessel function with the control parameter $g$ with $\Omega=10$ and $\mu=g/\Omega^2$. (b) Plot of the effective potential $V_{\mathrm{eff}}$ given by Eq.~(\ref{e7}) for three values of $g$ with $\beta=2$ and $\Omega=10$.}
\label{f2}
\end{figure}

Figure {\ref{f2}}(a) depicts the variation of $I_0(\mu)$ and $I_0(2\mu)$ with $g$ for $\Omega=10$. As $g$ increases from zero, the values of $I_0(\mu)$ and $I_0(2\mu)$ increase from the value $1$. $I_0(2\mu)$ increases more rapidly than $I_0(\mu)$ with $g$. Figure~{\ref{f2}}(b) shows the change in the shape of the effective potential with $g$. The location of the minimum of the potential $V_{\mathrm{eff}}(X)$ or the  $X$-component of the equilibrium point of the system (\ref{e6}) in absence of the periodic driving force is given by
\begin{equation}
 \label{e8}
  X^* = - \ln \left( \frac{I_0(\mu)}{I_0(2\mu)} \right).
\end{equation}
Because $I_0(2\mu)>I_0(\mu),~X^*$ is always $>0$ and it moves away from the origin as $g$ increases. Consequently, the depth $\Delta V_{\mathrm{eff}}=\mid V_{\mathrm{eff}}(X^*)\mid=\beta I_0^2(\mu)/2I_0(2\mu)$ decreases with increase in $g$ and the potential $V_{\mathrm{eff}}$ becomes more and more flat. Later, we point out an important consequence of this on the response amplitude $Q$.

A slow oscillation takes place about $X^*$. Therefore, for convenience, we introduce the change of variable $Y=X-X^*$ so that the slow oscillation occurs around $Y^*=0$. In terms of $Y$, Eq.~(\ref{e6}) becomes
\begin{subequations}
 \label{e9}
\begin{eqnarray}
  \ddot{Y} + d \dot{Y} + \omega_{\mathrm{r}}^2{\mathrm{e}}^{-Y} 
     \left( 1 - {\mathrm{e}}^{-Y} \right)
       =f \cos \omega t,
\end{eqnarray}
where
\begin{eqnarray}
  \omega_{\mathrm{r}}^2 = \beta \frac{I_0^2(\mu)}{I_0(2\mu)}.    
\end{eqnarray}
\end{subequations}
For $\vert f \vert \ll 1$ it is reasonable to assume that the amplitude of $Y$ is small so that we write series expansions for ${\mathrm{e}}^{-Y}$ and ${\mathrm{e}}^{-2Y}$ and neglect the nonlinear terms in $Y$. This results in the linear equation
\begin{equation}
 \label{e10}
  \ddot{Y} + d \dot{Y} + \omega_{\mathrm{r}}^2Y
       =f \cos \omega t.    
\end{equation}
$\omega_{\mathrm{r}}$ is the resonant frequency of oscillation of the slow motion. In the long time limit, the solution of Eq.~(\ref{e10}) is $Y=Qf\cos(\omega t+\phi)$ where
\begin{subequations}
 \label{e11}
\begin{eqnarray}
   Q = \frac{1}{\sqrt{S}}, \quad 
   S = \left( \omega_{\mathrm{r}}^2 - \omega^2 \right)^2
          + d^2 \omega^2,
\end{eqnarray}
and
\begin{eqnarray}
   \phi = \tan^{-1} \left( \frac{d\omega}{\omega^2
            - \omega_{\mathrm{r}}^2}\right),
\end{eqnarray}
\end{subequations}
where $Q$ is the response amplitude of the system (\ref{e2}) at the low-frequency $\omega$ of the input signal.

\subsection{Analysis of the vibrational resonance}
In order to verify the theoretical predictions to be obtained from the analysis of $Q$ given by Eq.~(\ref{e11}a), we calculate $Q$ from the numerical solution of Eq.~(\ref{e2}). We use the formula $Q=\sqrt{Q_{\mathrm{s}}^2+Q_{\mathrm{c}}^2}/f$ where [\ref{r3},\ref{r5},\ref{r13}]
\begin{subequations}
 \label{e12}
\begin{eqnarray}
   Q_{\mathrm{s}} 
     & = & \frac{2}{nT}\int_0^{nT}x(t)\sin\omega t~dt,\\
   Q_{\mathrm{c}}
     & = & \frac{2}{nT}\int_0^{nT}x(t)\cos\omega t~dt,
\end{eqnarray}
\end{subequations}
with $T= 2 \pi /\omega$ and $n$ is big enough, say $500$.

\begin{figure}[t]
\begin{center}
\includegraphics[width=0.85\linewidth]{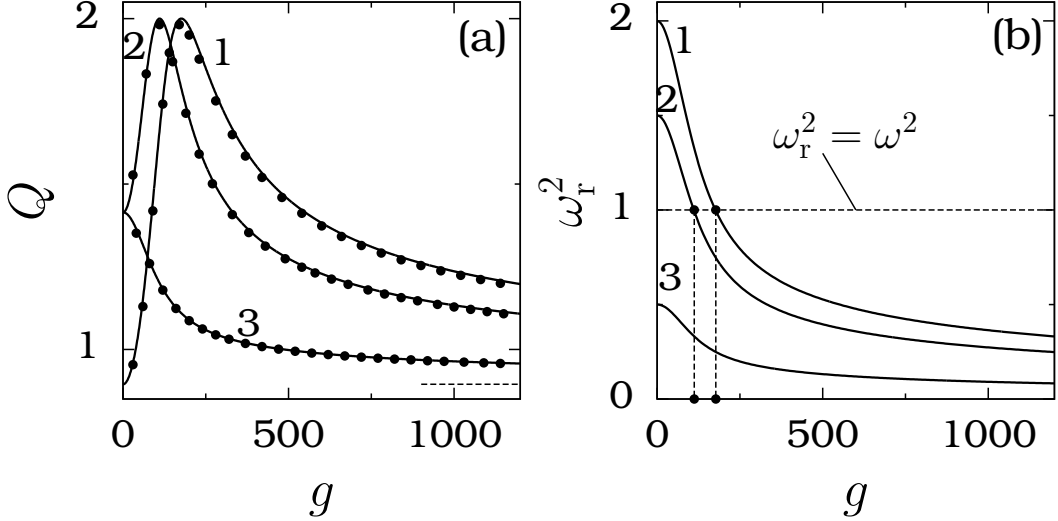}
\end{center}
\caption{(a) Response amplitude $Q$ versus the control parameter $g$ for the Morse oscillator for three values of $\beta$. The values of the parameters are $d=0.5,~f=0.1,~\omega=1$ and $\Omega=10$. The values of $\beta$ for the curves 1, 2 and 3 are $2,~1.5$ and $0.5$ respectively. The continuous curve and solid circles represent theoretically and numerically computed values of $Q$ respectively. The horizontal dashed line denotes the limiting value of $Q$, $Q_{\mathrm L}(g\rightarrow\infty)$. (b) Dependence of $\omega_{\mathrm{r}}^2$ with $g$ for the three values of $\beta$ used in the subplot (a). The solid circles on the $g$ axis denote the values of $g$ at which resonance occurs.}
\label{f3}
\end{figure}
     
The value of a control parameter at which resonance occurs ($Q$ becomes maximum) corresponds to the minimum of the function $S$ given by Eq.~(\ref{e11}a). The following are the key results of analysis of the theoretical expression of the response amplitude $Q$.
\begin{itemize}
\item 
For a fixed value of $g$ when $\omega$ is varied resonance occurs when $\omega=\omega_{_\mathrm{VR}}$ given by (obtained from $dS/d\omega=0$)
\begin{equation}
  \label{e13}
  \omega_{_\mathrm{VR}} = \sqrt{\omega_{\mathrm{r}}^2
     - \frac{d^2}{2}}, 
        \quad \omega_{\mathrm{r}}^2>\frac{d^2}{2}.
\end{equation}
\item
When $g$ is varied the condition for resonance is $dS/dg=4\omega_{\mathrm{r}}\omega_{\mathrm{rg}}(\omega_{\mathrm{r}}^2-\omega^2)=0$. Because $I_0$ is always $>0$ and increases monotonically with $g$ we have $\omega_{\mathrm{r}}\neq 0$ and $\omega_{\mathrm{rg}}=d\omega_{\mathrm{r}}/dg\neq0$ and hence the resonance condition is $\omega_{\mathrm{r}}^2=\omega^2$, that is, $\beta I_0^2(g/\Omega^2)/I_0(2g/\Omega)=\omega^2$. Resonance occurs whenever the resonant frequency $\omega_{\mathrm{r}}$ matches with the low-frequency $\omega$ of the periodic force.
\item
Figures \ref{f3}(a) and (b) show the variation of the response amplitude $Q$ and $\omega_{\mathrm{r}}^2$ respectively with $g$ for three fixed values of $\beta$ and $d=0.5,~f=0.1,~\omega=1$ and $\Omega=10$. The theoretical value of $Q$ is in very good agreement with $Q$ obtained from the numerical solution of Eq.~(\ref{e2}). In Fig.~\ref{f3}(a) for both $\beta=2$ and $1.5$ as $g$ increases from zero the value of $Q$ increases monotonically, it reaches a maximum at $g=g_{_\mathrm{VR}}$ and then decreases. For $\beta=2$ theoretical and numerical values of $g_{_\mathrm{VR}}$ are $176$ and $173$ respectively. In Fig.~\ref{f3}(b) for both $\beta=2$ and $1.5$ at $g=g_{_\mathrm{VR}}$  (indicated by solid circles),  $\omega_{\mathrm{r}}^2=\omega^2$.
\item
At $g=0$, $I_0(\mu)=I_0(2\mu)=1$ and $\omega_{\mathrm{r}}^2(g=0)=\beta$. As $g$ increases $I_0(\mu)$ and $I_0(2\mu)$ increases rapidly with $I_0(2\mu)$ growing faster than $I_0(\mu)$ as shown in Fig.~\ref{f2}(a). Consequently, $\omega_{\mathrm{r}}^2$ decreases rapidly from the value of $\beta$ for a while and then decays to zero slowly. The maximum value of $\omega_{\mathrm{r}}^2$ is $\beta$ and this happens at $g=0$ For $\beta<\omega^2$, $\omega_{\mathrm{r}}^2$ is always less than $\omega^2$, that is, $\omega_{\mathrm{r}}^2\neq\omega^2$ for $g>0$ implying no resonance. This is shown in Fig.~\ref{f3}(a) for $\beta=0.5$ and $\omega =1$ for which $Q$ decreases continuously with $g$.

\begin{figure}[t]
\begin{center}
\includegraphics[width=0.7\linewidth]{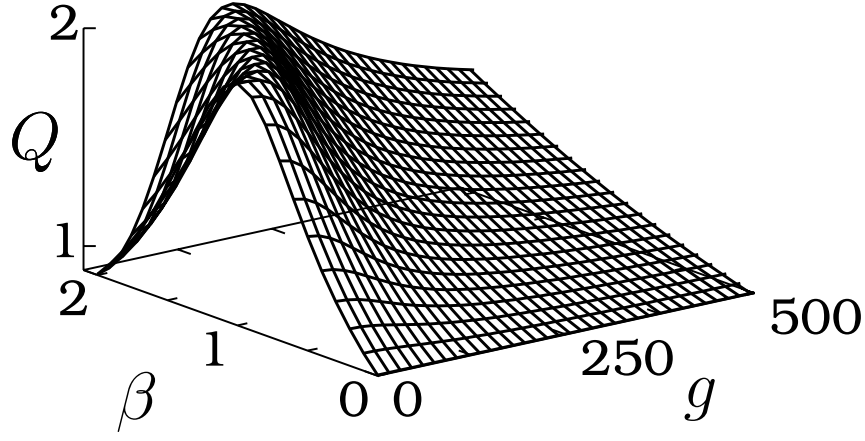}
\end{center}
\caption{Three-dimensional plot of the theoretically computed response amplitude $Q$ as a function of $\beta$ and $g$.}
\label{f4}
\end{figure}

\item
Resonance is possible only for $\beta>\omega^2$. In this case, as $g$ increases from zero, $\omega_{\mathrm{r}}^2$ decreases and becomes $\omega^2$ at only one value. Hence, there is only one resonance. This is shown in Fig.~\ref{f3}(a) for $\beta=1.5$ and $2$ with $\omega =1$. Figure \ref{f4} presents the dependence of $Q$ on $\beta$ and $g$ for $\omega=1$. Resonance is seen only for $\beta >\omega^2(=1)$.
\item
At resonance $\omega_{\mathrm{r}}^2=\omega^2$ and hence $Q_{\mathrm{max}}=1/(d\omega )$ and it depends only on $d$ and $\omega$. The response amplitude at resonance is independent of the parameters $\beta,~g$ and $\Omega$.

\begin{figure}[t]
\begin{center}
\includegraphics[width=0.55\linewidth]{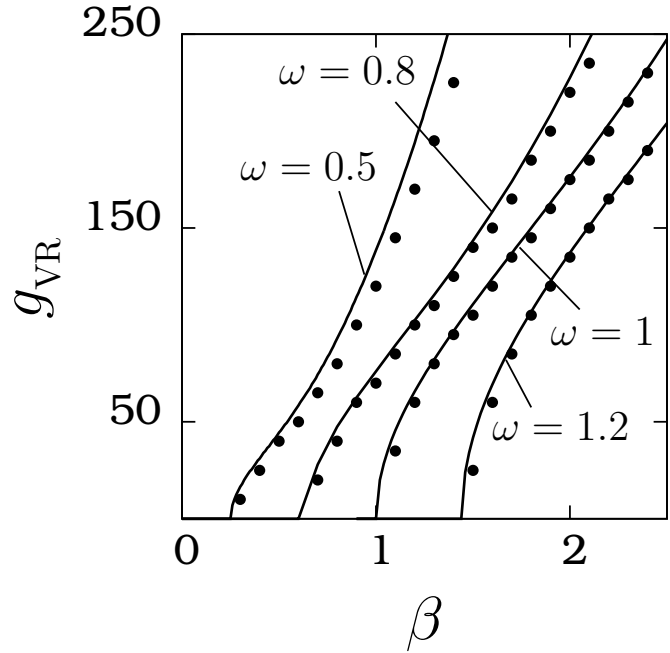}
\end{center}
\caption{Variation of theoretically predicted (continuous curve) and numerically computed (solid circles) $g_{_\mathrm{VR}}$ with the parameter $\beta$ for a few fixed values of the parameter $\omega$. The value of $\Omega$ is $10\omega$.}
\label{f5}
\end{figure}

\item
An analytical expression for $g_{_\mathrm{VR}}$ (at which $Q$ becomes maximum) is difficult to obtain because $\omega_{\mathrm{r}}^2$ is a complicated function of $g$. However, $g_{_\mathrm{VR}}$ can be calculated from the resonance curve. It depends on $\Omega,~\omega$ and $\beta $ and independent of $d$ and $f$. $Q$ decreases with increase in the value of $d$. In Fig.~\ref{f5} we plot the theoretical $g_{_\mathrm{VR}}$ and numerically computed $g_{_\mathrm{VR}}$ versus the parameter $\beta $ for a few fixed values of $\omega $. $g_{_\mathrm{VR}}$ increases with increase in the value of $\beta $.
\item
For very large values of $g,~\omega_{\mathrm{r}}^2\rightarrow 0$ and $Q$ approaches the limiting value $Q_{\mathrm{L}}$ given by
\begin{equation}
  \label{e14}
  Q_{_\mathrm{L}} (g\rightarrow\infty)
    = \frac{1}{\omega\sqrt{\omega^2+d^2}} .
\end{equation}
That is, $Q$ does not decay to zero but approaches the above limiting value (see Fig.~\ref{f3}(a)). This limiting value depends only on the parameters $\omega $ and $d$ (note that $Q_{\mathrm{max}}=1/(d\omega)$). The point is that when $\omega_{\mathrm{r}}^2\rightarrow0$, Eq.~(\ref{e9}) becomes the damped free particle driven by the periodic force whose solution is
\begin{equation}
 \label{e15}
    Y(t) = Q_{_\mathrm{L}} f \cos(\omega t+\Phi),
            \quad \Phi = \tan^{-1}(d/\omega) .  
\end{equation}
For the Duffing and quintic oscillators [\ref{r3},\ref{r5},\ref{r11},\ref{r25}] $V(x)$ (as well as the effective potential)$\rightarrow \infty$ as $x\rightarrow \pm \infty$. In this case the resonant frequency diverges after a few oscillations and thus $Q$ decays to zero for large values of $g$.
\end{itemize}
\section{Resonance with a square-wave signal}
In this section we show that vibrational resonance can be realized in the Morse oscillator when the external signal is a square-wave and the theoretical analysis employed in the previous section for sinusoidal force can be applied to this case also.

In the system (\ref{e2}) in place of $F_1=f\cos\omega t+g\cos\Omega t$ we consider three other forms of external force given by
\begin{eqnarray}
 \label{e16}
      F_2(t)
   & = & f \cos\omega t + g\,{\mathrm{sgn}}
           (\cos\Omega t), \\  
\label{e17}
     F_3(t) 
   & = & f\,{\mathrm{sgn}}(\cos\omega t)
            +g\cos\Omega t,\\
\label{e18}
     F_4(t) 
   & = & f\,{\mathrm{sgn}} (\cos\omega t)
          + g\,{\mathrm{sgn}}(\cos\Omega t),
\end{eqnarray}
where sgn$\vert u \vert$ denotes sign of $u$. In the theoretical analysis we use a Fourier series expansion for the square-wave signal.  For the Morse oscillator driven by the periodic force $F_2(t)$ the solution of slow motion is $Y_2=Q_2f\cos(\omega t+\phi_2)$  where
\begin{subequations}
\label{e19}
\begin{eqnarray}
     Q_2 = \frac{1}{\sqrt{(\omega_{\mathrm{r,2}}^{2}
            -\omega^{2})^2+d^{2}\omega^{2}}},
\end{eqnarray}
where
\begin{eqnarray}
    \omega_{\mathrm{r,2}}^2
    & =  & \frac{\beta U^{2}(\mu)}{U(2\mu)}, 
          \quad U(\mu) = 
           \frac{1}{2\pi}\int_0^{2\pi}{\mathrm{e}}^{-\psi_2}
             ~d\tau, \\
    \psi_2
    & = & \frac{4\mu}{\pi}\sum _{n=0}^{\infty}\frac{(-1)^{n+1}
           \cos(2n+1)\tau}{(2n+1)^3}.
\end{eqnarray}
\end{subequations}
For the case of the system driven by the force $F_3$ the solution $Y_3$ is 
\begin{subequations}  
\label{e20}
\begin{equation}
   Y_3 = \frac{4f}{\pi}\sum _{n=0}^{\infty}
          \frac{(-1)^n}{(2n+1)}\frac{\cos[(2n+1)\omega t
          +\phi_3 ]}{ \left[ \left( \omega_{\mathrm{r,3}}^2
         -(2n+1)^2\omega ^2 \right)^2 +(2n+1)^2d^2\omega ^2 
          \right]^{\frac{1}{2}}},
\end{equation}
where
\begin{equation}
  \omega_{\mathrm{r,3}}^2
    = \frac{\beta I_0^2(\mu)}{I_0(2\mu)}.
\end{equation}
\end{subequations}
The solution $Y_3$ has frequencies $\omega$ and odd integer multiples of it. The response amplitude $Q_3$ corresponding to the fundamental frequency $\omega$ is 
\begin{equation}
\label{e21}
   Q_3 = \frac{4}{\pi}\frac{1}{\sqrt{ \left( 
           \omega_{{\mathrm{r,3}}}^{2}-\omega^2 \right)^2
             + d^2 \omega^2}}.  
\end{equation} 
When the biharmonic force is chosen as $F_4$ given by Eq.(\ref{e18}) then the expression for the solution $Y_4$ is the same as $Y_3$ except that now $\omega_{{\mathrm{r,3}}}^2$ is replaced by $\omega_{{\mathrm{r,4}}}^2$ where 
\begin{equation}
\label{e22}
  \omega _{\mathrm{r,4}}^{2}= \frac{\beta U^{2}(\mu)}{U(2\mu)}.    
\end{equation}
Then
\begin{equation}
  \label{e23}
   Q_4 = \frac{4}{\pi}\frac{1}{\sqrt{ \left( 
         \omega_{{\mathrm{r,4}}}^2-\omega^2 \right)^2
           + d^2 \omega^2}} \;.
\end{equation}
%
 
\begin{figure}[t]
\begin{center}
\includegraphics[width=0.55\linewidth]{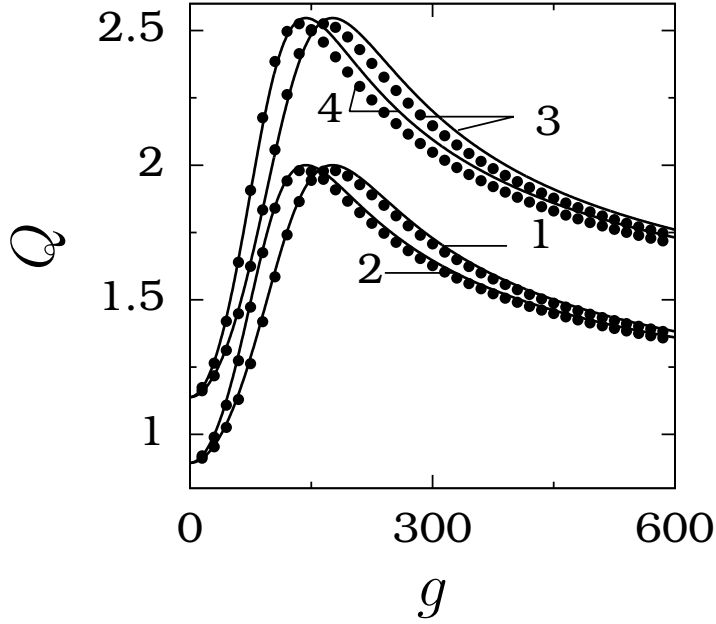}
\end{center}
\caption{Variation of the response amplitude $Q$ with $g$ for four different types of biharmonic forces. The continuous curve and the solid circles represent theoretically calculated $Q$ and numerically calculated $Q$ respectively. The biharmonic forces corresponding to the curves and solid circles marked as 1, 2, 3 and 4 are $f\cos\omega t+g\cos\Omega t$, $F_2,~F_3$ and $F_4$ respectively.}
\label{f6}
\end{figure}

Figure {\ref{f6}} shows both theoretically determined $Q$ and numerically computed $Q$ versus $g$ with different input signals where $d=0.5,~\beta=2,~f=0.1,~\omega=1$ and $\Omega=10$. In this figure, apart from close agreement of theoretical $Q$ with numerical $Q$, we notice that $Q_{{\mathrm{1,max}}}=Q_{{\mathrm{2,max}}}$, $Q_{{\mathrm{3,max}}}=Q_{{\mathrm{4,max}}}$, $Q_{{\mathrm{1,L}}}=Q_{{\mathrm{2,L}}}$ (limiting values of $Q$ in the limit of $g\rightarrow \infty$) and $Q_{{\mathrm{3,L}}}=Q_{{\mathrm{4,L}}}$. These results can be accounted from the theoretical expressions of $Q_i$'s. The resonance condition  $\omega_{{\mathrm{r}},i}^2=\omega^2$, $i=1,~2,~3,~4$ in the expression of $Q_i$'s gives
\begin{subequations}
\label{e24}
\begin{eqnarray}
  Q_{{\mathrm{1,max}}}
    & = & Q_{{\mathrm{2,max}}} = \frac{1}{d\omega},\\
  Q_{{\mathrm{3,max}}} 
    & = & Q_{{\mathrm{4,max}}}  
            = \frac{4}{\pi d\omega}
            =\frac{4}{\pi}Q_{{\mathrm{1,max}}},
\end{eqnarray}
\end{subequations}
that is, $Q_{{\mathrm{3,max}}}$ and $Q_{{\mathrm{4,max}}}$ are $4 / \pi\approx  1.27324$ times of $Q_{{\mathrm{1,max}}}$. For the parametric values used in our analysis, $Q_{{\mathrm{1,max}}}=2$ and hence $Q_{{\mathrm{3,max}}}$ and $Q_{{\mathrm{4,max}}}$ are $2.54648$ as is the case in Fig.~{\ref{f6}}. For sufficiently large values of $g$, we have $\omega_{{\mathrm{r}},i}^2\approx 0$  and hence
\begin{equation}
\label{e25}
   Q_{\mathrm{1,L}} = Q_{\mathrm{2,L}}
      = \frac{1}{\omega\sqrt{\omega^{2}+d^2}},
         \quad Q_{\mathrm{3,L}}
         = Q_{\mathrm{4,L}}
         = \frac{4}{\pi}Q_{\mathrm{1,L}} .
\end{equation}
Furthermore, because of $\omega_{\mathrm{r,1}}^{2}=\omega_{\mathrm{r,3}}^{2}$ and $\omega_{\mathrm{r,2}}^{2}=\omega_{\mathrm{r,4}}^{2}$ we find that 
\begin{equation}
\label{e26}
  Q_3 = \frac{4}{\pi}Q_{1}, \quad Q_4 = \frac{4}{\pi}Q_2.   
\end{equation}
That is, the response amplitude at the frequency $\omega$ when the input signal is a square-wave with fundamental frequency being $\omega$ is $4 / \pi$ times that of the signal $f \cos \omega t$. Numerical results in Fig.~{\ref{f6}} confirms all the above theoretical predictions.

Our analysis shows that the form of the low- and high-frequency forces need not be identical. For any arbitrary force containing a component with frequency $\omega$, enhancement of the amplitude of the output signal at the frequency $\omega$ can be achieved by using another arbitrary force containing a frequency $\Omega\gg \omega$.  When the force involved are simple periodic function of $t$ then the theoretical analysis of vibrational resonance is very much feasible.
\section{Quantum mechanical Morse oscillator}
In the previous two sections we focused our analysis on vibrational resonance in the classical Morse oscillator.  In the present section we are concerned with the quantum mechanical Morse oscillator in the presence of the  biharmonic external field $W(t)=F \cos \omega t + g \cos \Omega t$ with $\Omega \gg \omega$.  When a quantum mechanical system is subjected to a time-dependent external field the system undergoes transition between the energy eigenstates.  Therefore, we are interested in knowing the probability of finding the system in an $f$th state at time $t$.

\begin{figure}[t]
\begin{center}
\includegraphics[width=0.75\linewidth]{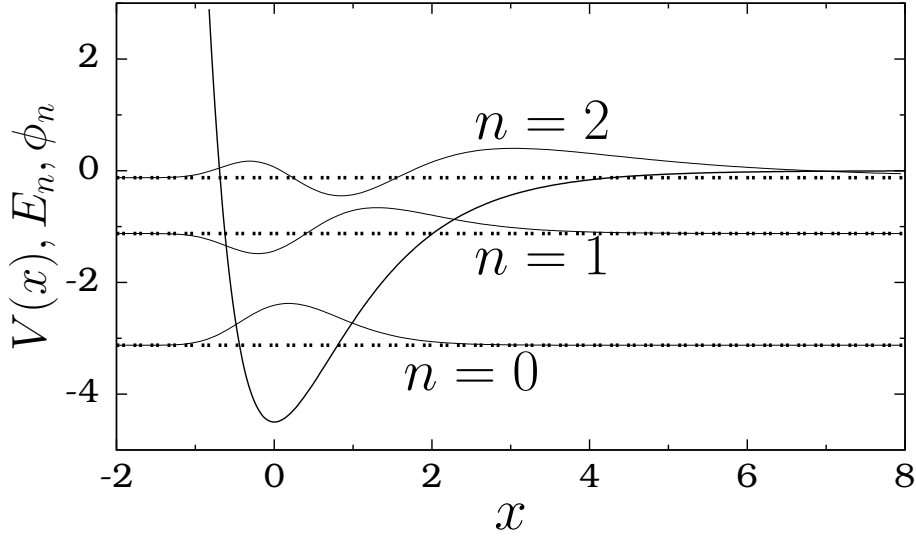}
\end{center}
\caption{Bound state energy eigenvalues and eigenfunctions of the Morse oscillator for $\beta =9$.  The dashed lines are the energy eigenvalues. The potential $V(x)$ is also shown.}
\label{f7}
\end{figure}

The unperturbed Hamiltonian of the system is $H_0=p_x^2/(2m)+V(x)$ where $V(x)$ is given by Eq.~(\ref{e1}).  The unperturbed system $H_0\phi_n=E_n \phi_n$ is exactly solvable and the eigenfunctions and energy eigenvalues are given by [\ref{r38}-\ref{r40}] 
\begin{subequations}
\label{e27}
\begin{eqnarray}
   \phi_n  = N_n z^{\lambda -n-1/2} \,{\mathrm{e}}^{-z/2} L_n^k (z), 
\end{eqnarray}
where
\begin{eqnarray}
    z & = & 2 \lambda {\mathrm{e}}^{-x}, \;\; 
              \lambda^2 = \frac{m \beta}{\hbar^2}, \;\;
              N_n = \left( \frac{k \,n!}{(2 \lambda -n-1)!} 
              \right)^{1/2}, \\
     k & = & 2 \lambda-2n-1, \;\;L_n^k (z)  = \frac{z^{-k} 
                   {\mathrm{e}}^z}{n!}
                    \frac{d^n}{dz^n} {\mathrm{e}}^{-z}
                    z^{n+k}              
\end{eqnarray}
\end{subequations}
and 
\begin{equation}
\label{e28}
   E_n = - \frac{\hbar^2}{2m} 
           \left( \lambda -n - \frac{1}{2} \right)^2, \quad
           n=0,1,\cdots \; {\mathrm{and}} \; n < \lambda -\frac{1}{2}.
\end{equation}
In Eq.~(\ref{e27}c) $L_n^k(z)$ are the generalized Laguerre polynomials.  The Morse oscillator has a finite number of bound states and the number of bound states can be controlled by the parameter $\beta$.  Setting the values of $\hbar$ and $m$ as unity for convenience and $\beta=9$ we obtain
\begin{equation}
\label{e29}
    E_0 = -3.125, \;  E_1  =  -1.125, \; E_2 = -0.125. 
\end{equation}
There are only three bound states.  Figure {\ref{f7}} shows the energy eigenvalues and the eigenfunctions for $\beta =9$.

In the presence of the external field the Hamiltonian of the system is $H=H_0+ \epsilon H_1$, where $H_1=xW(t)$.  The time-dependent Schr\"odinger  equation for the system is
\begin{equation}
\label{e30}
    {\mathrm{i}} \hbar \frac{\partial \psi}{\partial t}
     = H \psi ,
\end{equation}
where $\psi(x,t)$ is the wave function of the perturbed system.  We write
\begin{equation}
\label{e31}
   \psi (x,t) = \sum_n a_n(t) \phi_n(x) 
                {\mathrm{e}}^{-{\mathrm{i}}E_n t/\hbar} .
\end{equation}
The probability of finding the system in the state $n$ is $P_n(t) = \vert a_n(t) \vert ^2$, $\sum_n \vert a_n(t) \vert^2 =1$. 

To determine $a_n(t)$ we apply the standard time-dependent perturbation theory [\ref{r41}].  Suppose the external field is switched-on at $t=0$ and switched-off at $t=T$, that is, the external field is applied during a finite time interval $T$.  Assume that the system is initially in a state $i$ with the eigenfunction $\phi_i$.  Then at $t=0$ the probability of finding the system in the state $i$ is $1$ and the probability of finding the system in the other states is $0$: $a_n(0)=\delta_{ni}$.  Due to the applied field  the system can make a  transition from the state $i$ to an another state  after the time $T$.  Once the perturbation is switched-off the system settles down to a stationary state and denote this final state as $f$.  

To determine $a_n(t)$ we substitute Eq.~(\ref{e31}), $a_f=a_f^{(0)}+\epsilon a_f^{(1)} + \cdots$ and equate the terms containing various powers of $\epsilon$ to $0$.  Up to first-order in $\epsilon$, after some algebra,  we obtain $a_f^{(0)}=\delta_{fi}$ and
\begin{subequations}
\label{e32}
\begin{eqnarray}
    a_f^{(1)}(T) = \frac{C_{fi}}{2\hbar} s, 
\end{eqnarray}
where 
\begin{eqnarray}
    s & = & F(r_{1+}+r_{1-})+g(r_{2+}+r_{2-}), \\
    r_{1\pm} & = & \frac{1-{\mathrm{e}}^{{\mathrm{i}} (\omega_{fi}
                 \pm \omega) T }}{\omega_{fi}\pm \omega}, \quad 
    r_{2\pm}  = \frac{1-{\mathrm{e}}^{{\mathrm{i}} (\omega_{fi}
                 \pm \Omega) T }}{\omega_{fi}\pm \Omega},
\end{eqnarray}
\begin{eqnarray}
  \omega_{fi} = (E_f-E_i)/\hbar, \quad 
   C_{fi} = \int_{-\infty}^{\infty} \phi_f^* x \phi_i 
              \, {\mathrm{d}} x .
\end{eqnarray}
\end{subequations}
The transition probability for $i$th state to $f$th state is given by $P_{fi}(T) = \left| \delta_{fi}+ \epsilon a_{f}^{(1)}(T) \right| ^2$.  In $a_f^{(1)}(T)$ the term $s$ alone depends on the parameters $F$, $\omega$, $g$ and $\Omega$ of the external field and $T$.  Therefore, we study the variation of the quantity $\vert s \vert^2$ with the parameters of the external field.

\begin{figure}[t]
\begin{center}
\includegraphics[width=0.8\linewidth]{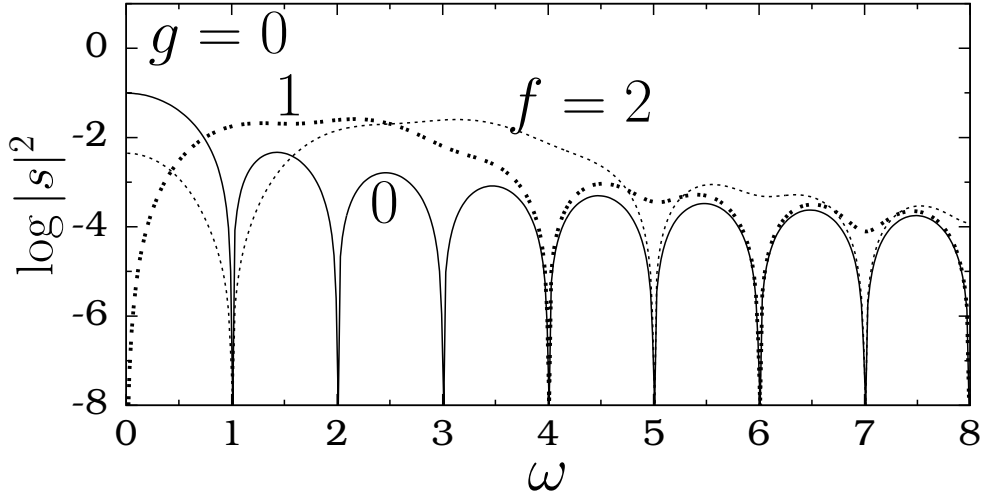}
\end{center}
\caption{$\log \vert s \vert^2$ versus $\omega$ for the states $f=0$, $1$ and $2$ for $\beta=9$, $F=0.05$, $g=0$, $T=\pi$ and $i=0$.}
\label{f8}
\end{figure}

We fix $\beta=9$, $F=0.05$, $T=\pi$ and assume that the system is initially in the ground state ($i=0$).  Figure \ref{f8} shows the variation of $\log \vert s \vert^2$ with $\omega$ for $g=0$. The first-order correction to transition probability displays a sequence of resonance peaks with decreasing amplitude.  
The following results are evident from Eqs.~(\ref{e32}b-c) and Fig.~\ref{f8}.  The quantity $s$ consists of only $r_{1+}$ and $r_{1-}$.  We have $\omega_{00}=0$, $\omega_{10}=2$ and $\omega_{20} = 3$.  Consequently, for the states $f=0$, $1$ and $2$ the quantity $r_{1+}$ can be neglected when $\omega \approx 0$, $2$ and $3$ respectively because the denominator in $r_{1-}$ is $\approx 0$.  Thus, the first-order transition probability for the states $f=0$, $1$ and $2$ becomes maximum at $\omega=0$, $2$ and $3$ respectively.  For $\omega \approx 0$ the values of $\vert s \vert^2$ for $f=0$, $1$ and $2$ are $\approx 4F^2 \pi^2$, $0$ and $16F^2/9$ respectively.  The first-order transition probability for the state $f=1$ is $\approx 0$.  For the $f=0$ state $\vert s \vert^2=0$ when $\omega=1,2,\cdots$ and it becomes maximum when $\omega=n+\frac{1}{2}$, $n=1,2,\cdots$.  In the case of $f=1$ at $\omega=4,6,\cdots$ the numerator in both $r_{1+}$ and $r_{1-}$ becomes zero and thus the quantity $\vert s \vert^2$ is minimum.  For $f=2$,  at odd integer values of $\omega$ except at $\omega=3$, $r_{1 \pm}=0$  and hence $\vert s \vert^2$ becomes minimum.

\begin{figure}[t]
\begin{center}
\includegraphics[width=0.73\linewidth]{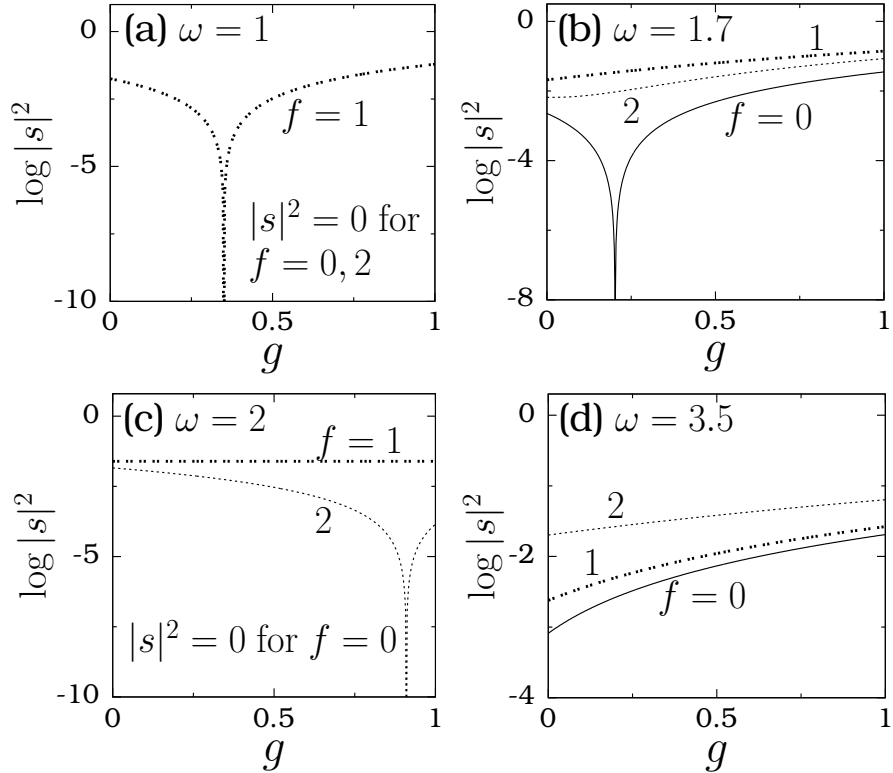}
\end{center}
\caption{Variation of $\vert s \vert^2$ (in $\log$ scale) with the amplitude $g$  of the high-frequency  external field for four fixed values of the frequency  $\omega$ of the low-frequency external field.  The value of $\Omega$ is fixed as $5 \omega$. }
\label{f9}
\end{figure}

Next, we include the high-frequency field and vary its amplitude $g$ and the frequency $\Omega$. Figure \ref{f9} presents the results for a few fixed values of $\omega$ with $\Omega=5 \omega$.  When $\omega=1$, in Eq.~(\ref{e32}b), $r_{1+}+r_{1-}=0$ and $r_{2+}+r_{2-}=0$ for $f=0$ and $2$ and hence the increase of $g$ has no effect on $\vert s \vert^2$.  For the $f=1$ state we find $s=(8/3)F-(8/21)g$.  As $g$ increases $\vert s \vert ^2$ decreases from $(8/3)F$, becomes $0$  at $g=7F(=0.35)$ and then increases with further increase in $g$ as shown in Fig.~\ref{f9}(a).  The first-order transition probability of the state $f=1$ exhibits anti-resonance at $g=0.35$.  Anti-resonance can be realizable for other states also for appropriate choices of $\omega$.  For example, in Figs.~\ref{f9}(b) and (c) corresponding to $\omega=1.7$ and $2$ respectively we can clearly notice anti-resonance for $f=0$ state and $f=2$ state.   In Fig.~\ref{f9}(d) where $\omega=3.5$  the quantity $\vert s \vert^2$ increases monotonically with the control parameter $g$ for all the three states.  

\begin{figure}[!h]
\begin{center}
\includegraphics[width=0.73\linewidth]{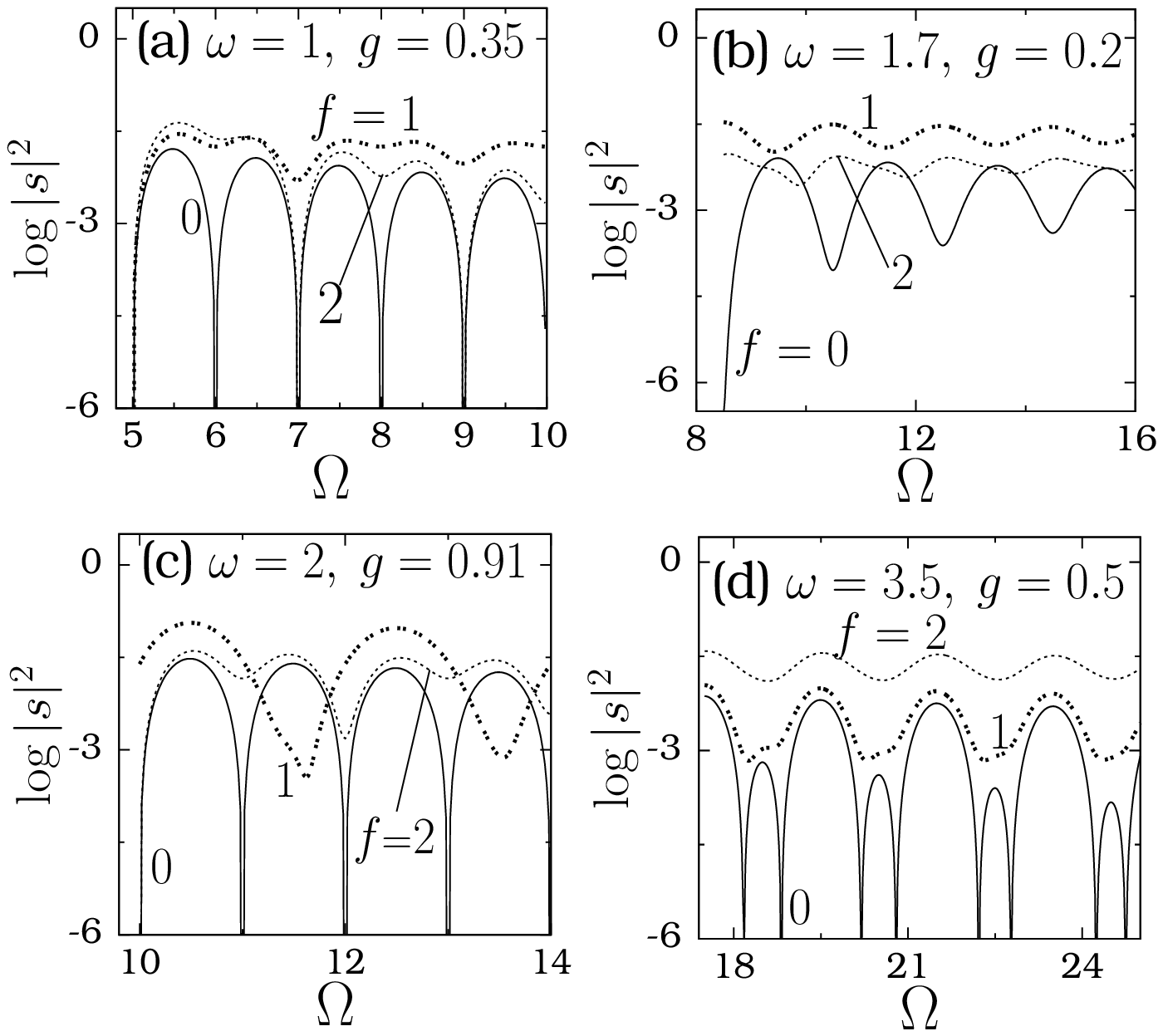}
\end{center}
\caption{$\log \vert s \vert^2$ versus $\Omega$ (the frequency of the high-frequency external field) for four sets of values of $\omega$ and $g$.}
\label{f10}
\end{figure}

Because $r_{2+}$ and $r_{2-}$ (given by Eq.~(\ref{e32}c)) contain terms which are sinusoidal functions of $\Omega$ the first-order transition probability can exhibit a sequence of resonance peaks when $\Omega$ is varied for fixed values of the other parameters.  This is shown in Fig.~\ref{f10} for four sets of values of $\omega$ and $g$.  In all the cases $\Omega$ is varied from $5\omega$. In Figs.~\ref{f10}(a-c) $\vert s \vert^2=0$ at the starting value of $\Omega$ for the $f=0$ state.  $\vert s \vert^2 \ne 0$ for wide ranges of values of $\Omega$.  Transition probability of all the states show a sequence of resonance peaks.  In Fig.~\ref{f10}(d) $\vert s \vert^2$ of $f=0$ state is close to its values of the other two states.  However, as $\Omega$ increases $\vert s \vert^2$ of the $f=1$ and $f=2$ states oscillate but $\ne 0$.  In contrast to this, $\vert s \vert^2 $ of the $f=0$ state becomes $0$ at certain values of $\Omega$.

It is noteworthy to compare the effect of  high-frequency external force in the classical and quantum mechanical Morse oscillators.  A classical nonlinear system can exhibit a variety of dynamics when a control parameter is varied.  However, in the Morse oscillator for the choice $ \vert f \vert \ll 1$ and $\Omega \gg \omega$ when the amplitude $g$ of the high-frequency force is varied the system is found to show only periodic motion with period $T=2 \pi/\omega$.  The response amplitude of the motion exhibits a single resonance when the control parameter $g$ or $\omega$ or $\Omega$ is varied.  Resonance occurs whenever the resonant frequency $\omega_{{\mathrm{r}}}$ (Eq.~(\ref{e9}b)) matches with the frequency $\omega$.  In the case of the quantum mechanical Morse oscillator we have considered  the simple case of switching-on the external field at $t=0$ and switching-off it at $t=T$.  In the absence of the high-frequency field the first-order transition probability $P_{fi}$ shows a sequence of resonances with decreasing  amplitude when the parameter $\omega$ is increased. The dominant resonance occurs at $\omega=\omega_{fi}$.   Resonance is not observed when the amplitude $g$ of the high-frequency field is varied.  However, anti-resonance of $P_{fi}$ takes place for certain values of $\omega$.  Multiple resonance of $P_{fi}$ occurs when the frequency $\Omega$ of the high-frequency field is varied. 

\section{Conclusion}
We have reported our investigation on high-frequency periodic force induced resonance at the low-frequency component of the output of a single oscillator. Using a perturbation theory an analytical expression for the response amplitude is obtained. Interestingly, from the analytical expression of $Q$ we are able to derive various features of the vibrational resonance and its mechanisms. The occurrence of the resonance depends on the parameter $\beta$ and $\omega$ while the values of the response amplitude at resonance and for large values of $g$ depend only on the damping coefficient $d$ and the low-frequency $\omega$. In the system the limiting values of $Q$ is nonzero because for sufficiently large values of $g,~\omega_{\mathrm r}^{2}\approx 0$. From the analytical expression of $Q$ given by Eq.(\ref{e11}a) we note that $Q$ becomes a nonzero constant if $\omega_{\mathrm{r}}^{2}\approx$ a constant. The theory used in our present analysis can be applied to the other different forms of input signal than $\cos\omega t$ and $\sin\omega t$. We have demonstrated its applicability for the case of square-wave form. All the theoretical results are well supported by the numerical simulation.  We have considered the quantum version of the Morse oscillator in the presence of the biharmonic external field.  Interestingly, a high-frequency external field is found to induce resonance and anti-resonance on the transition probability for the transition from an $i$th state to an $f$th state.  If the transition probability $P_{fi}$ of a state is weak in the presence of a harmonic external field with a particular frequency $\omega$, then it can be enhanced by an another external field of relatively high-frequency.  The dominance of a state can be changed by the high-frequency external field.  That is, $P_{fi}$ can be controlled by a second harmonic external field.

\section*{{\bf{Acknowledgments}}}
KA acknowledges the support from University Grants Commission (UGC), India
in the form of UGC-Rajiv Gandhi National Fellowship.  Financial support from the Spanish Ministry of Science and Innovation under Project No. FIS2009-09898 is acknowledged by MAFS.

\section*{References}
\markright{References}
\renewcommand{\labelenumi}{[\theenumi]}
\renewcommand{\theenumi}{\arabic{enumi}}
\begin{enumerate}
\item
\label{r1}
L~Gammaitoni, P~H\"{a}nggi, P~Jung and F~Marchesoni, {\emph{Rev. Mod. Phys.}} {\bf{70}}, 223 (1998)
\item
\label{r2}
M D~McDonnell, N G~Stocks, C E M~Pearce and D~Abbott, {\emph{Stochastic resonance}} (Cambridge University Press, Cambridge, 2008)
\item
\label{r3}
P S~Landa  and P V E~McClintock, {\emph{J. Phys. A: Math. Gen.}}  {\bf{33}}, L433 (2000)
\item
\label{r4}
M~Gittermann,{\emph{ J. Phys. A: Math. Gen.} {\bf{34}}}, L355 (2001)
\item
\label{r5}
I I~Blechman and P S~Landa,  {\emph{Int. J. Nonlin. Mech.}} {\bf{39}}, 421 (2004)
\item
\label{r6}
V~In, A~Kho, J D~Neff, A~Palacios, P~Loghini and B K~Meadows, {\emph{Phys. Rev. Lett.}} {\bf{91}}, 244101 (2003)
\item
\label{r7}
V~In, A R~Bulsara, A~Palacios, P~Loghini and A~Kho, {\emph{Phys. Rev. E.}} {\bf{72}}, 045104(R) (2005)
\item
\label{r8}
B J~Breen, A B~Doud, J R~Grimm, A H~Tanasse, S J~Janasse, J F Lindner and K J~Maxted, {\emph{Phys. Rev. E}} {\bf{83}}, 037601 (2011)
\item
\label{r9}
E Ippen, J Lindner and W L~Ditto, {\emph{J. Stat. Phys.}} {\bf{70}}, 437 (1993)
\item
\label{r10}
S Zambrano, J M~Casado and M A F~Sanjuan,
{\emph{Phys. Lett. A}} {\bf{366}}, 428 (2007)
\item
\label{r11}
S~Jeyakumari, V~Chinnathambi, S~Rajasekar and M A F~Sanjuan, {\emph{Phys. Rev. E}} {\bf{80}}, 046608 (2009)
\item
\label{r12}
S~Rajasekar, K~Abirami and M A F~Sanjuan, {\emph{Chaos}} {\bf{21}}, 033106 (2011)
\item
\label{r13}
E~Ullner, A~Zaikin, J~Garcia-Ojalvo, R~Bascones and J~Kurths, {\emph{Phys. Lett. A}} {\bf{312}}, 348 (2003)
\item
\label{r14}
H~Yu, J~Wang, C~Liu, B~Deng and X~Wei, {\emph{Chaos}} {\bf{21}}, 043101 (2011)
\item
\label{r15}
V M~Gandhimathi, S~Rajasekar and J~Kurths, {\emph{Phys. Lett. A}} {\bf360}, 279 (2006)
\item
\label{r16}
B~Deng, J~Wang and X~Wei, {\emph{Chaos}} {\bf{19}}, 013117 (2009)
\item
\label{r17}
B~Deng, J~Wang, X~Wei, K M~Tsang and W L~Chan, {\emph{Chaos}} {\bf{19}}, 013113 (2010)
\item
\label{r18}
V N~Chizhevsky, E~Smeu and G~Giacomelli, {\emph{Phys. Rev. Lett.}} {\bf{91}}, 220602 (2003) 
\item
\label{r19}
V N~Chizhevsky and G~Giacomelli, {\emph{Phys. Rev. E}} {\bf{70}}, 062101 (2004)
\item
\label{r20}
V N~Chizhevsky and G~Giacomelli, {\emph{Phys. Rev. E}} {\bf{73}}, 022103 (2006)
\item
\label{r21}
J P~Baltanas, L~Lopez, I I~Blechman, P S~Landa, A~Zaikin, J~Kurths and M A F~Sanjuan, {\emph{Phys. Rev. E}} {\bf{67}}, 066119 (2003) 
\item
\label{r22}
C~Yao, Y~Liu and M~Zhan, {\emph{Phys. Rev. E}} {\bf{83}}, 061122 (2011)
\item
\label{r23}
C~Yao and M~Zhan, {\emph{Phys. Rev. E}} {\bf{81}}, 061129 (2010)
\item
\label{r24}
S~Rajasekar, J~Used, A~Wagemakers and M A F~Sanjuan, {\emph{Commun. Nonlinear Sci. Numer. Simulat.}} {\bf{17}}, 3435 (2012)
\item
\label{r25}
S~Jeyakumari, V~Chinnathambi, S~Rajasekar and M A F~Sanjuan, {\emph{Chaos}} {\bf{21}}, 275 (2011)
\item
\label{r26}
J H~Yang and X B~Liu, {\emph{J.Phys. A: Math. Theor.}} {\bf{43}}, 122001 (2010)
\item
\label{r27}
J H~Yang and X B~Liu, {\emph{Chaos}} {\bf{20}}, 033124 (2010)
\item
\label{r28}
C~Jeevarathinam, S~Rajasekar and M A F~Sanjuan, {\emph{Phys. Rev. E}} {\bf{83}}, 066205 (2011)
\item
\label{r29}
J H~Yang and X B~Liu, {\emph{Phys. Scr.}} {\bf{83}}, 065008 (2011)
\item
\label{r30}
A~Ichiki, Y~Tadokoro and M I~Dykman, {\emph{Phys. Rev. E.}} {\bf{85}} 031107 (2012)
\item
\label{r31}   
J R~Ackerhalt and P W~Milonni, {\emph{Phys. Rev. A}} {\bf{34}}, 1211 (1986)
\item
\label{r32}   
M E~Goggin and P W~Milonni, {\emph{Phys. Rev. A}} {\bf{37}}, 796 (1988)
\item
\label{r33}   
D~Beigie and S~Wiggins, {\emph{Phys. Rev. A}} {\bf{45}}, 4803 (1992)
\item
\label{r34}   
A~Memboeuf and S~Aubry, {\emph{Physica D}} {\bf{207}}, 1 (2005)
\item
\label{r35}   
W~Knob and W~Lauterborn, {\emph{J. Chem. Phys.}} {\bf{93}}, 3950 (1990)
\item
\label{r36}   
Z~Jing, J~Deng and J~Yang, {\emph{Chaos, Solitons and Fractals}} {\bf{35}}, 486 (2008)
\item
\label{r37}
K T~Tang, {\emph{Mathematical methods for engineers and scientists: Fourier analysis, partial differential equations and variational models}} (Springer, Berlin, 2007) pp191
\item
\label{r38}
A~Frank, R~Lemus, M~Carvajal, C~Jung and E~Ziemniak, {\emph{Chem. Phys. Lett.}} {\bf{308}}, 91 (1999)
\item
\label{r39}
R~Lemus and A~Frank, {\emph{Chem. Phys. Lett.}} {\bf{349}},  471 (2001)
\item
\label{r40}
R~Lemus, {\emph{J. Mol. Spectrosc.}} {\bf{225}}, 73 (2004)
\item
\label{r41}
L I~Schiff, {\emph{Quantum mechanics}}, (McGraw-Hill, New York, 1968)
\end{enumerate}

\end{document}